\documentclass[12pt,a4paper]{article}
\usepackage[utf8]{inputenc}
\usepackage[left=2.75cm,right=2.75cm,top=3cm,bottom=3cm]{geometry}

\usepackage{amsmath}
\usepackage{amsfonts}
\usepackage{amssymb}
\usepackage{amsthm}
\usepackage{bbm}
\usepackage{bm}
\usepackage{mathtools}
\usepackage{graphicx}
\usepackage{caption}
\usepackage[flushleft]{threeparttable}
\usepackage[dvipsnames]{xcolor}
\usepackage{booktabs}
\usepackage{rotating}
\usepackage{multirow}
\usepackage{longtable}
\usepackage{setspace}
\usepackage{relsize}
\usepackage{graphicx,subcaption}
\usepackage{pdflscape} 
\usepackage{xurl}
\usepackage{ragged2e}
\usepackage{float}
\usepackage[colorlinks=true]{hyperref}
 \hypersetup{citecolor=MidnightBlue, anchorcolor=Black, urlcolor=MidnightBlue, linkcolor=MidnightBlue}
\usepackage[nottoc]{tocbibind}
\usepackage{comment}
\usepackage{placeins}

\usepackage[backend=biber, style=apa, bibstyle=authoryear, maxcitenames=2, maxbibnames=6, giveninits=true, dashed = false, sorting=nyt, date=year, eprint = false, url=false, doi=false, isbn=false]{biblatex} 
\newcommand\fnote[1]{\captionsetup{font = footnotesize}\caption*{#1}}

\title{\Large Quantitative Tools for Time Series Analysis in Natural Language Processing: A Practitioners Guide\footnote{I am grateful for helpful discussions with Melanie Althage, Max Grossmann, Anselm K\"usters, and exceptionelly valuable coding advice from Achim Zeileis, especially on the details of the `strucchange' package. \emph{All errors are my own.} 
}
}

\author{W.~Benedikt Schmal\thanks{Ilmenau University of Technology, Department of Economic Sciences and Media, Economic Theory Group, Ehrenbergstraße 29, 98693 Ilmenau, Germany.\\
Email: \href{mailto:wbschmal@gmail.com}{wbschmal@gmail.com}. ORCiD: \href{https://orcid.org/0000-0003-2400-2468}{0000-0003-2400-2468}.}\\
\normalsize Ilmenau University of Technology\\
\normalsize DICE, Heinrich Heine University\\
\normalsize MSI, KU Leuven University
}

\date{May 1, 2024}
\begin{document}
\maketitle
\thispagestyle{empty}

\begin{onehalfspace}
\begin{center}
\vspace{4mm}

\textbf{Abstract}
\end{center}

Natural language processing tools have become frequently used in social sciences such as economics, political science, and sociology. Many publications apply topic modeling to elicit latent topics in text corpora and their development over time. Here, most publications rely on visual inspections and draw inference on changes, structural breaks, and developments over time. We suggest using univariate time series econometrics to introduce more quantitative rigor that can strengthen the analyses. In particular, we discuss the econometric topics of non-stationarity as well as structural breaks. This paper serves as a comprehensive practitioners guide to provide researchers in the social and life sciences as well as the humanities with concise advice on how to implement econometric time series methods to thoroughly investigate topic prevalences over time. We provide coding advice for the statistical software R throughout the paper. The application of the discussed tools to a sample dataset completes the analysis.\\

\noindent \emph{Keywords:} NLP, natural language processing, text analysis time series, econometrics, non-stationarity, structural breaks\\
\noindent \emph{JEL Codes:} A20, C22, C45, C87
\end{onehalfspace}
\newpage

\begin{doublespace}
\section{Introduction}
\label{sec.int}

Topic modeling is a computational method that transforms qualitative text into quantitative data that can be measured with a plethora of statistical methods. By that, it has become an invaluable tool for systematically investigating large amounts of text data and its contents. The foundations were laid by \textcite{Blei.2003} when they developed the Latent Dirichlet Allocation algorithm to structure text data into unspecified, latent topics. Since then, many refinements, variations, and extensions have been developed to capture a variety of underlying data-generating processes properly \parencite[See][for a review]{Vayansky.2020}. 

All the described tools are not new or innovative but quite established econometric methods that have been known for years, if not decades. Furthermore, macroeconomics, the economics subfield in which time series analysis played an essential role, shifted towards other methods and dismissed traditional univariate time series econometrics. Nevertheless, these techniques still need to be realized in many academic disciplines outside economics. Furthermore, work by \textcite{Laureate.2023} finds that researchers using topic models for investigations of social media texts do not use them in the best way. While the present paper does not address the application of topic models directly, it indirectly explains how advanced time series methods can be used with topic probabilities computed based on topic modeling.

Topic modeling has become an essential tool in social and life sciences, and there already exists many practitioners' guides for the use of topic models per se \parencite[see, e.g.,][]{Karl.2015, Fu.2019, Isoaho.2019} or the appropriate number of topics to choose, which has to be defined exogenously \parencite[see, e.g.,][]{Greene.2014, Weston.2023}.\footnote{Note that these are examples and, by far, not exhaustive.} Nevertheless, many developments over time continue to be measured solely by visual inspection and interpretation of time series. It applies particularly strongly to research in the digital humanities that often encompasses extensive periods in which the analyzed text documents had been published \parencite[for example,][]{Wehrheim.2019, Wehrheim.2020, Kusters.2020, Kusters.2023}. One recent example that tries to change this is the work by \textcite{Schmal.2023}. He first elicits latent topics using a structural topic model \parencite{Roberts.2019} and extracts the expected probabilities for each topic and each paper. In the second step, he categorizes the topics and aggregates the topic probabilities by category and year to conduct time series analytics. The present paper ties in with this work and wants to provide the curious reader a manual to follow the steps of \textcite{Schmal.2023} when conducting time series econometrics. There already exists work aiming at providing non-economists with practical advice on how to study time series econometrically \parencite[see, e.g.,][]{Jebb.2017}. Closest to this paper is the work by \textcite{Zeileis.2003}, who describe how to detect structural breaks in time series using the R software package. The present paper comprehensively describes non-stationarity and how to detect it. Furthermore, the specific link between time series econometrics and topic modeling remains a research desiderat. 

In general, the present paper will introduce practitioners in social science research to these methods. It may also be helpful for students who want to enhance their essays or theses with quantitative rigor. This paper cannot be exhaustive but serves as an invitation to read further and dig deeper to foster more thorough quantitative research in the social sciences and digital humanities. Throughout the paper, we provide immediate coding suggestions using the statistical software \verb|R|, published and maintained under an open-source license and freely available. Throughout the paper, we refer several times to econometrics textbooks and encourage the reader to consult these resources when conducting their own research.\footnote{If one wants to use econometric tools more broadly in R, among others, \textcite{Kleiber.2008} provide a textbook covering that.}

\begin{figure}[htbp]
\centering
\includegraphics[width = .8\linewidth]{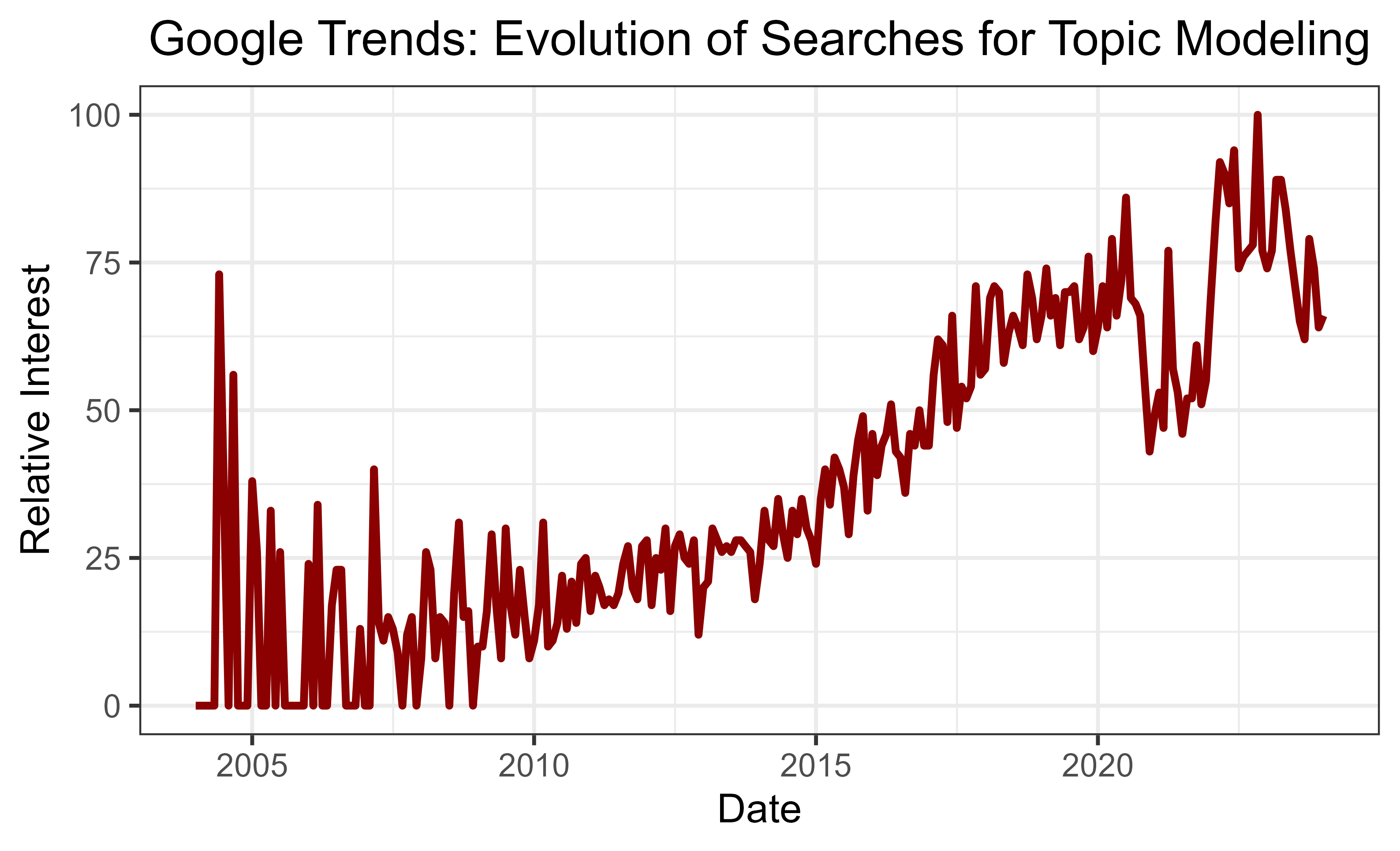}
\fnote{Data received from Google Trends for the phrase ``topic modeling'' on January 2024 using the URL \url{https://trends.google.de/trends/explore?date=all&q=topic\%20modeling&hl=de}. Data for search requests worldwide. Note that Google Trends computes interest in a search term \emph{relative} to the date with the highest interest, coded as 100. Based on that, interest for all other dates is computed relative to the highest number. Raw numbers are not available from Google.}
\caption{\normalsize Evolution of Google searches for `topic modeling' over time}
\label{fig.gtrend}
\end{figure}

As an illustration, we investigate the time series for the evolution of Google searches for ``topic modeling'' from January 2004 until January 2024.\footnote{The dataset is available on Zenodo, see \textcite{Schmal.2024zen1}.} Figure \ref{fig.gtrend} displays the time series as a line plot. Visual inspection displayed high interest and high fluctuation until 2010. Since then, there seems to have been a steady upward trend that has only been interrupted in more recent years, namely between mid-2020 and early 2022. Since then, it has stabilized at a high level and decreased again in recent months.\\

In the following sections, I will discuss mainly two critical quantitative analyses of time series data. First, in Section \ref{ssec.stat}, I will elaborate on the question of stationarity or non-stationarity within a time series. It is an essential tool for detecting various trends in a time series. Secondly, in Section \ref{ssec.struc}, I will address the question of the presence of a structural break within observations, which is often only solved by visual inspection. Applying statistical tools can overcome this. Furthermore, the endogenous search for such structural changes will be explained. Section \ref{sec.conc} briefly concludes.
\pagebreak

\section{Stationarity}
\label{ssec.stat}
\vspace{-5mm}
\subsection{Statistical foundations}

Stationarity, or non-stationarity, is a core concept for understanding time series data. The former implies that the distribution of the outcome variable does not depend on time. Put differently, stationarity requires a constant mean and variance for the outcome variable across time. In addition, the covariance between two outcome observations must depend only on its temporal distance (i.e., three years), but not \emph{which} three years. Again, in different terms, stationarity implies a reversion to the mean on the one hand and no growth in terms of fluctuations over time \parencite{Hill.2008}. Spoken in applied terms, stationarity within a time series implies that the effect of a shock fades out over time. If it triggers a persistent change, for example, due to a change in preferences of the individuals, this should be reflected by non-stationarity within the data. 

Non-stationarity implies that at least one of the three stated criteria is violated. In general, there exist three types of non-stationarity \parencite[see, e.g.,][]{Hill.2008}, and for computed topic prevalences, one should understand what kind of non-stationarity could be present. We speak of a `random walk' if an outcome in period $t$ depends on the outcome in a previous period $t-x$ as, in this case, there is no mean reversion of a clear dependence of today's values on past values. Note that a random walk alone is stochastic and does not require a precise movement in one direction (i.e., upward or downward). 

In addition, such a movement could be captured by a drift and/or a deterministic trend. Without getting into too much detail, a drift implies a fixed term added in every time series period. For example, we start in $t_0$ with value $y_0=1$. In $t_1$, we have $y_1 = y_0 + 0.5 + \epsilon_1$, where $y_0 =1$, $epsilon_1$ is the random error term, and 0.5 is the additive drift. In our example, we get $y_1 = 1.5+\epsilon_1$. Accordingly, in $y_2$, we have $y_2 = 2+\epsilon_1 + \epsilon_2$. Thus, the drift of 0.5 slowly accumulates into one fixed direction (upwards in our case as $0.5>0$). The error term $\epsilon_t$ varies over time and may be negative or positive. An additional deterministic trend, however, is understood to be time-dependent. In this case, we do not add (for example) 0.5 in each period but $0.5t$, which implies an added term of 0.5 in $t_1$, but 2 in $t_4$ -- instead of 0.5 in $t_4$ in case of a drift.

\subsection{Non-stationarity in topic modeling}

Before elaborating further on (non-)stationarity and how to test for it, we should briefly think about the `$1-p$' problem when studying probabilities \parencite[see, e.g.,][]{Schmal.2023a, Schmal.2023}. Topic probabilities, by construction, add up to one in any case. Thus, for a set of $N$ topics within a topic model, a growth in topic $t_i$ necessarily implies a decrease in $\sum^N_{j\neq i}t_j\:\forall j \in \{1,...,N\}$. Due to that fact, non-stationarity, particularly in the mean, is likely to be an issue one should address when applying topic models. Because if one topic (non-stationary) grows over time, other topics are likely to shrink and vice versa. 

Measuring and discussing the non-stationarity of computed topic probabilities over time is not just a technical exercise, but a meaningful endeavor. In topic modeling, absolute values are often the focus of discussion due to the open question of which cut-off values should be considered appropriate when discussing topic prevalences. However, it is a priori unclear what absolute level particular expected probabilities should reach to be considered important, specific, or relevant. This is where non-stationary changes over time come in. They measure relative differences, regardless of the absolute probability level of a certain topic. This approach is immune to the problem of arbitrariness in the choice of discrete absolute cut-off values, making it a more nuanced and impactful method for investigating topic dynamics. 

\subsection{Testing for non-stationarity}

Statisticians have developed plenty of testing methods for stationarity. I want to discuss two established methods applicable to many cases of suspected stationarity in topic prevalence, but we encourage the readers to extend their knowledge. We begin with the Augmented Dickey-Fuller (ADF) test \parencite{Dickey.1979, Said.1984} that poses as a null hypothesis the presence of a so-called `unit root,' which points to the non-stationarity of the time series. The alternative hypothesis is the absence of a unit root or else stationarity. The other test method discussed later on is the KPSS test.\footnote{Please be aware that there exist further testing procedures, among others, the test by \textcite{Phillips.1988}, the one by \textcite{Ng.2001}, as well as the Elliot-Rothenberg-Stock Test \parencite{Elliott.1996}.} Before moving on, one has to be aware of the particular testing procedure of the ADF test. As the null hypothesis captures non-stationarity, failure to reject it implies that one cannot rule out the presence of a unit root, but it is not clear evidence that one exists \parencite[see on this issue also][]{Verbeek.2008}. The ADF test is a generalization of the original testing procedure by \textcite{Dickey.1979} and allows for three specifications, namely non-stationarity without drift or trend, which implies a random walk, non-stationarity with drift, and non-stationarity with both drift and trend. To do so in \verb|R|, we can use the \verb|aTSA| package developed by Debin Qiu.\footnote{See \url{https://cran.r-project.org/package=aTSA} for details.}
\begin{footnotesize}
\begin{verbatim}
library(aTSA)
ts <- data$value
aTSA::adf.test(ts)
\end{verbatim}
\end{footnotesize}
The command is relatively parsimonious and provides many results that need further interpretation. In addition to the three types, it also offers test statistics and (adjusted) p-values for several lag lengths. Lag implies an autoregressive process, i.e., the dependence of an outcome value $y$ in period $t$ on past values $t-x$. The lag defines this $x$ further. Statistical information criteria, rules of thumb, or theoretical reasoning can determine the optimal lag length. 

\begin{figure}[htbp]
\centering
\includegraphics[width = .8\linewidth]{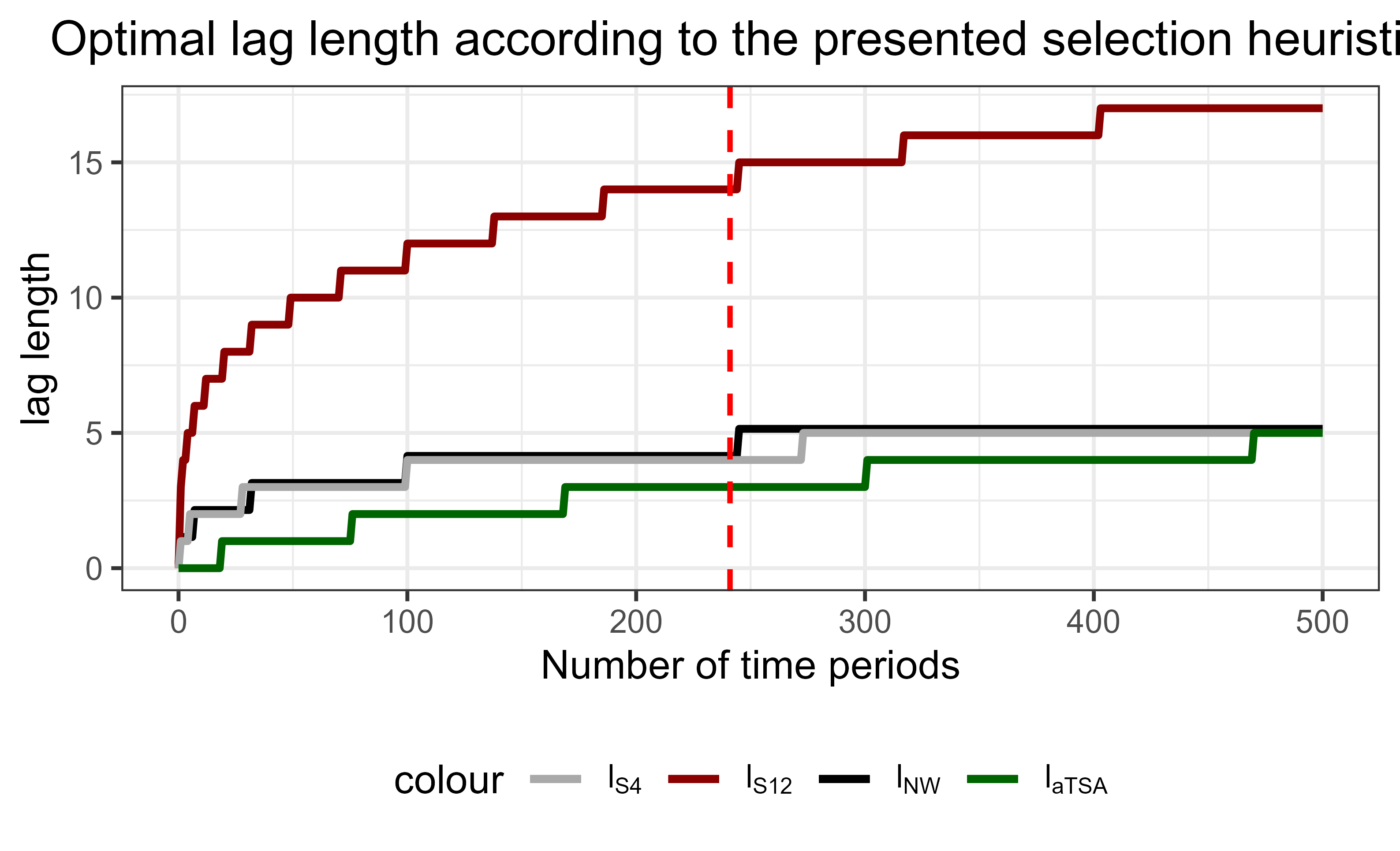}
\caption{\normalsize Optimal lag length by optimality criterion and length of the time series}
\label{fig.lag_length}
\end{figure}

Eventually, one has to decide which lag level to use in the ADF test. Here, the researcher must again determine which data-generating process underlies their particular data. One common rule of thumb is the rule by \textcite{Schwert.1989}. He suggests to version for the choice of the lag length, namely $l_{S4}$ and $l_{S12}$, where $l_{S4} = \lfloor \left(4\left(\frac{T}{100}\right)^{\frac{1}{4}}\right)\rfloor$ and $l_{S12} = \lfloor \left(12\left(\frac{T}{100}\right)^{\frac{1}{4}}\right)\rfloor$. One can see that only the factor differs for the two options. $\lfloor\:\rfloor$ symbol a floor function, i.e., one has to choose the lowest integer value of a computed result. Say we have $l=4.3$, then $\lfloor4.3\rfloor=4$ holds. \textcite{Newey.1994} suggest in their optimal lag length computation $l = \lfloor \left(4\left(\frac{T}{100}\right)^{\frac{2}{9}}\right)\rfloor$ as first step. One can already see that this has to be similar to $l_{S4}$ as the exponential terms $\frac{1}{4}\approx\frac{2}{9}$ are similar. Figure \ref{fig.lag_length} demonstrates graphically for $T \in [0,500]$ observations that, indeed $l_{S4}\approx l_{NW}$, while $l_{S12}$ is much higher. 

While the \verb|aTSA| package for the ADF test leaves it to the researcher to determine, the test procedure for the KPSS test comes with a pre-selection of the optimal lag length, which might be another indicator for the researcher which lag to choose for the ADF test. The KPSS test -- an abbreviation for its developers Kwiatkowski, Phillips, Schmidt, and Shin -- is another test for non-stationarity in a time series \parencite{Kwiatkowski.1992}. Compared to the ADF test, its primary difference and advantage are that it poses stationarity as the null hypothesis and the existence of a unit root, i.e., non-stationarity, as the alternative hypothesis. It is easy to compute with \verb|aTSA|:
\begin{footnotesize}
\begin{verbatim}
ts <- data$value
aTSA::kpss.test(ts, lag.short = TRUE)
\end{verbatim}
\end{footnotesize}
The \verb|lag.short| option is set to \verb|TRUE| by default and refers to the formula proposed by \textcite{Schwert.1989}. Hence, if the researcher wants to rely on that specification, they can ignore the second part of the previously presented command. As before, in the ADF test, the KPSS test provides us with three different types of non-stationarity that are investigated: Neither drift nor trend, only a drift, or both drift and trend. Ideally, the ADF and the KPSS test lead to similar, if not equivalent, results regarding the presence of a unit root. However, as always with real-life data, its characteristics often tend to be more opaque than econometric simulations in a controlled environment. 

That said, we turn towards our time series example using the Google Trends statistics for search requests for `topic modeling.'

\subsection{The `Google Trends' Example}

Visually investigating the time series shown in Figure \ref{fig.gtrend}, after a phase of intense fluctuation until late 2009, a clear upward trend becomes visible. Hence, topic modeling seems to have gained interest a couple of years after the computational foundations had been laid. It is reasonable since new methods need time to be disseminated across researchers, especially across different academic disciplines that are often more or less isolated from each other, so new methods are not established at the same time or speed. However, the upward trend between 2010 and 2020 hints at a unit root with drift as the increase appears to have a somewhat linear functional form. 

Thus, we compute both the ADF and the KPSS test for our sample. In \verb|R|, we call our dataset received from Google Trends `data,' and the percentage points we call `values.'\footnote{The whole code is available on \href{https://github.com/schmalwb/topic_model_time_series}{GitHub}, including the data used. The curious researcher can follow every step and reproduce every plot of this paper.} 
\begin{footnotesize}
\begin{verbatim}
ts <- data$value
aTSA::adf.test(ts, nlag = 5)
Augmented Dickey-Fuller Test 
alternative: stationary 
Type 1: no drift no trend 
     lag      ADF p.value
[1,]   0 -2.09207  0.0378
[2,]   1 -1.11267  0.2811
[3,]   2 -0.45610  0.5126
[4,]   3 -0.17046  0.5948
[5,]   4  0.00362  0.6449
Type 2: with drift, no trend 
     lag   ADF p.value
[1,]   0 -4.23  0.0100
[2,]   1 -2.77  0.0678
[3,]   2 -1.96  0.3415
[4,]   3 -1.69  0.4509
[5,]   4 -1.54  0.5074
Type 3: with drift and trend 
     lag   ADF p.value
[1,]   0 -9.98  0.0100
[2,]   1 -6.87  0.0100
[3,]   2 -4.89  0.0100
[4,]   3 -4.25  0.0100
[5,]   4 -3.88  0.0153
Note: in fact, p.value = 0.01 means p.value <= 0.01
\end{verbatim}
\end{footnotesize}
The output of the ADF test looks like this. Note that \verb|nlag = 5| specifies the number of lag specifications, including zero lags. Thus, the present test provides test statistics and the related p values for $l \in [0,4]$. We study up to four lags as $l_{S4}$ and $l_{NW}$ suggest $l^*=4$ for $T=241$, the number of periods we have in our dataset, see also the red dashed line in Figure \ref{fig.lag_length} shown beforehand.

One can see that we cannot reject non-stationarity assuming neither drift nor trend nor non-stationarity assuming a drift on the 5\% level for all lag specifications except for $l=0$. In fact, there is strong evidence for a unit root except for a specification with both drift and trend. Here, we reject the null of non-stationarity for every lag specification. Put differently; if we expected both drift and a deterministic trend, the actual data is stationary relative to this expectation. Put differently, if we expect in some unrelated case a jump in the data by 100\% but only observe 20\%, we might consider the change as `flat.' However, if we expect only an increase of 15\%, the observed 20\% is, indeed, a notable increase.

Hence, the ADF test is strong evidence for non-stationarity with a drift. Let us see what the KPSS test suggests:
\begin{footnotesize}
\begin{verbatim}
ts <- data$value
aTSA::kpss.test(ts, lag.short = TRUE)
KPSS Unit Root Test 
alternative: nonstationary 
Type 1: no drift no trend 
 lag stat p.value
   3 2.45  0.0171
 Type 2: with drift, no trend 
 lag stat p.value
   3  2.7    0.01
 Type 1: with drift and trend 
 lag  stat p.value
   3 0.452    0.01
Note: p.value = 0.01 means p.value <= 0.01, p.value = 0.10 means p.value >= 0.10 
\end{verbatim}
\end{footnotesize}
The output of the KPSS test is much shorter as it only considers one lag specification, namely three lags. It is due to the set-up of the \verb|kpss.test()| command of the \verb|aTSA| package in \verb|R|. The author, Debin Qiu, specifies the optimal lag length as $l = \lfloor \left(3\frac{\sqrt{T}}{13}\right)\rfloor$ other than suggested by \textcite{Kwiatkowski.1992}. As one can draw from the dark green line shown in Figure \ref{fig.lag_length} above, the \verb|aTSA| specification tends towards a shorter lag length. Unfortunately, the \verb|kpss.test()| does not allow a manual correction of the lag length to be used in the testing procedure. 

Turning towards the findings of the KPSS test for our Google Trends sample, we can reject the null hypothesis of stationarity for all three types of non-stationarity on the 5\% significance level. Hence, the KPSS test suggests even non-stationarity with both a drift and a trend. As stated beforehand, empirical data always comes with its specific foibles. Therefore, it is typical that ADF and KPSS tests may yield somewhat different results. However, the present analysis is strong evidence that (a) the time series is not stationary and (b) that non-stationarity with a drift seems to be present. In turn, it implies persistent growth of interest in topic modeling.

\section{Structural breaks}
\label{ssec.struc}
\vspace{-5mm}
\subsection{Statistical foundations}
Structural breaks are often not a sidenote but at the core of social science research. Much research discusses changes in response to surprising elections, the financial crisis, military interventions, or the COVID-19 pandemic. Thus, it is often a goal of plotting a time series to spot such a break. It can be visually supported by added vertical lines at a suspected point in time and, in addition, added horizontal lines that capture pre- and post-event averages of the variable of interest. These visual tools can already nudge a reader into seeing a change where not necessarily one exists in statistical terms. 

To provide more substantial evidence for a suspected shock or break, one can apply the \textcite{Chow.1960} test. Essentially, the test splits a dataset into two at a point chosen by the researcher. Then, it computes two separate regressions and compares whether the regression parameters significantly differ. If so, the null hypothesis of no break can be rejected in favor of the alternative hypothesis that a structural break exists. 

In \verb|R|, the Chow test can be executed with the \verb|sctest()| command, part of the \verb|strucchange| package \parencite{Zeileis.2002}. Here, within the brackets, one first defines the regression, which is in the given example just the outcome variable on the left that is regressed on the time variable, here `year.' \verb|type| specifies that a Chow test shall be used as there exist more than just this method to detect structural breaks. Lastly, \verb|point| defines the data point, e.g., in time, where or when the suspected shock happened. 
\begin{footnotesize}
\begin{verbatim}
library(strucchange)
strucchange::sctest(outcome ~ year, type = "Chow", point = x, data = data)
\end{verbatim}
\end{footnotesize}

It can be done with a quite parsimonious model, in which an outcome variable is regressed on a time variable. However, it can also be applied more broadly to more sophisticated specifications. 

\subsection{Detecting endogenous break points}

Up to now, we have only spoken about a known exogenous event, for which we would like to evaluate whether this shock is reflected in our chosen outcome variable. An alternative approach is the search for `endogenous breaks' as \textcite{Enders.2010} calls it. Here, one searches iteratively for a break without suspecting a specific date. To do that, one splits the used dataset into two subsets at every point in time and computes whether significant differences between the two regression models can be detected at this cut-off point. Theoretically, this can be done with a simple loop that iterates for a range of numbers through the breakpoint specifier in the above command. In reality, this comes with the challenge that the F-statistic used to detect a significant break is distorted as it incorrectly leads to too many rejections of the null of `no break.' Fortunately, we can address this again with the \verb|strucchange| package in \verb|R|. 

\textcite{Enders.2010} emphasizes that due to a decrease in statistical power, one should always have at least 10\% of all observations in each regression. Put differently, one should, at most, perform a 90/10 split. Of course, it is more of a rule of thumb. What matters is that the model parameters can be estimated with sufficient reliability. Depending on the frequency and the nature of the individual dataset, it might allow for a more extreme ratio or require a more conservative split as an upper threshold. In general, more even splits are preferable so that outliers unrelated to structural breaks do not distort the regressions and, by that, the results of the Chow test. It also implies that researchers should pay attention to the choice of their data sample in the sense that there should exist enough observations before and after a suspected shock one wants to study in some way. 

However, let us turn again towards the search for an endogenous break point. As said, we can, unfortunately, not just loop through the mentioned \verb|sctest()| command. However, \textcite{Zeileis.2002} did great work with \verb|strucchange| and provided respective tools to detect breakpoints. Assume a dataset with 100 observations. We need the lower and upper ten as minimum samples in this case. Therefore, we can iterate from obs \#11 - \#90 as shown below:
\begin{footnotesize}
\begin{verbatim}
result <- strucchange::Fstats(reg, from = 11, to = 90, data = data)
strucchange::sctest(result)
strucchange::plot(result)
strucchange::sctest(result)
strucchange::boundary(result, alpha = 0.05)
strucchange::boundary(result, alpha = 0.05, aveF = TRUE)
\end{verbatim}
\end{footnotesize}
Note that the regression model can be replaced by an element that stores it, i.e., you can set \verb|reg <- data$outcome ~ data$year| and insert only \verb|reg| into the commands above. Storing the results of the \verb|Fstats()| command can be helpful for the subsequent commands. Now, what are we doing in command 2 and 3? The \verb|sctest| command that processes the previous \verb|Fstats()| computes an $F$ test statistic and a corresponding p-value\footnote{Less relevant to know but highly important for the underlying statistics: This p-value has been adjusted for the multiple hypothesis tests conducted in this method as suggested by \textcite{Hansen.1997}.} that tests whether there exists a structural break within the range of potential breakpoints that have been entered into the command previously (here 11-89). 

To determine where the structural break lies, one can compute boundary values for the F statistic that show thresholds for certain confidence intervals. For example, the command \verb|strucchange::boundary(result, alpha = 0.05)| computes the critical F value that must be exceeded for a structural break with 95\% confidence. The \verb|boundary| command allows for two threshold specifications, the `supremum F' (the default specification of the \verb|R| command, which, therefore, does not need to be separately specified)  and the `average F' condition \parencite[see for the statistical background][]{Andrews.1994}, which is enforced by \verb|aveF = TRUE| in \verb|R|. For the former, more conservative criterion, it matters whether the highest F value for the presence of a structural break becomes too large not to reject the null hypothesis of no break. For the latter, the criterion is based on the average F statistic, which, by definition, also includes smaller F values, lowering the threshold's overall size. This, in turn, may lead to more detected structural breaks. 

A crucial caveat for practitioners is not to fall prey to hindsight bias: The breakpoint measures are computed based on an ex-post analysis and based on the time window the researcher provides. Therefore, one \emph{must not} infer from such a computation from which point in time onward a structural break occurs. It is wrong hindsight bias to claim to know when -- considering later developments -- a structural break occurred for the first time, falsely considering data points that did not exist yet when the alleged break happened.

\subsection{The `Google Trends' Example}

Returning to the evolution of searches for `topic modeling' using Google's search engine, running a test for a structural break based on the comparison of two separate regressions does not make sense for breakpoints until 2010 as the high level of fluctuations would inevitably lead to approximately to an insignificant coefficient for time no matter which date one would use. In contrast, there would be a significant positive estimator for the second regression, which always leads to the mechanical detection of a structural break. This artifact serves as a valuable illustration that statistical analyses are an important \emph{complement} to more qualitative approaches, but certainly not a replacement. 

Ignoring the years 2004 until 2010, it is easy to spot visual breaks between mid-2020 and the end of 2022, where one can observe a u-shaped pattern that interrupts the steady increase in searches for `topic modeling.' Thus, it would be reasonable to identify a structural break in 2020 -- the COVID-19 pandemic easily comes to mind as it affected researcher productivity \parencite{Deryugina.2021} and productivity in general \parencite{Fischer.2022}. The term itself is technical and, therefore, unlikely to be driven by non-scientific search requests. 

\begin{figure}
\centering
\includegraphics[width = .8\linewidth]{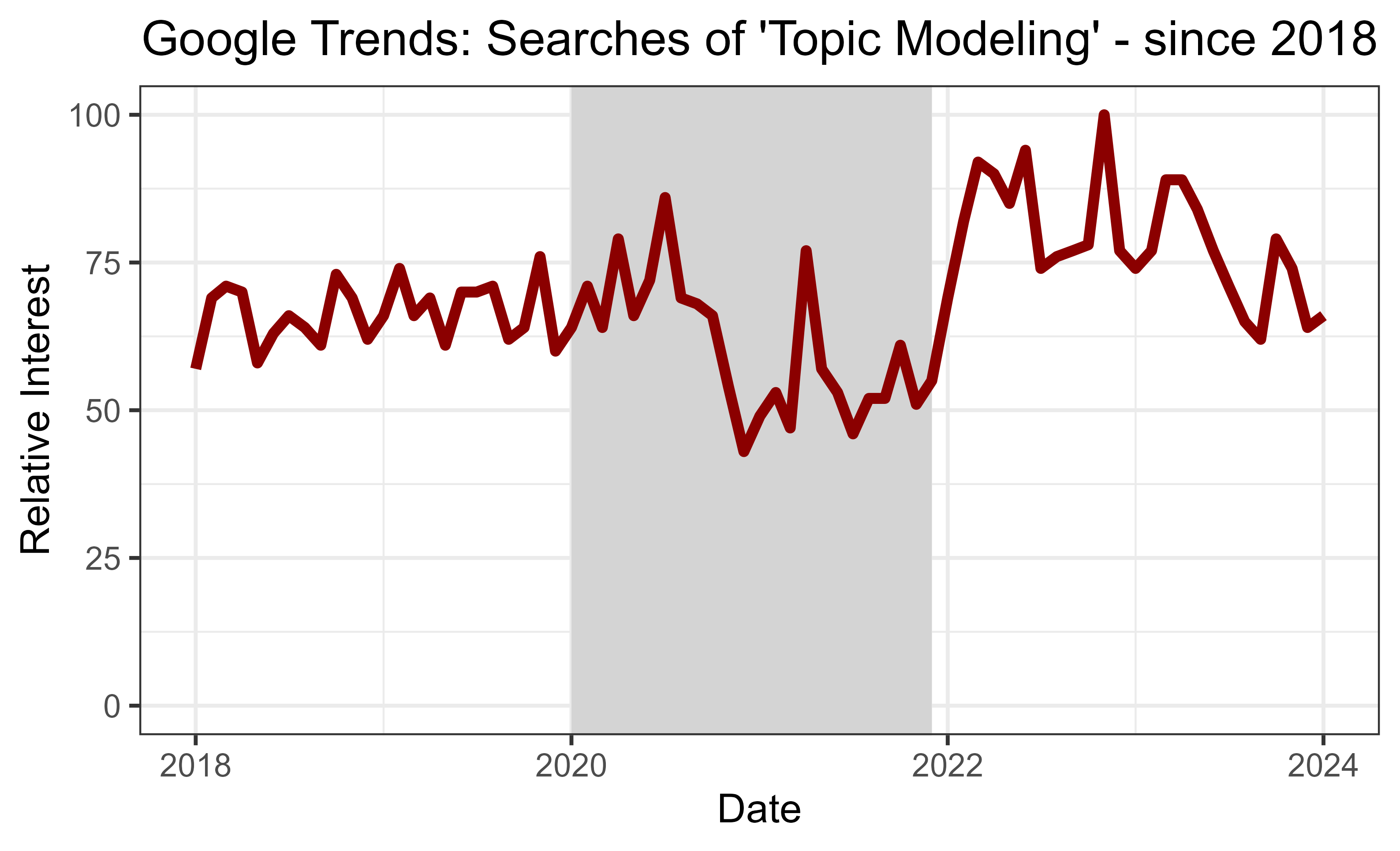}
\caption{\normalsize Evolution of Google searches for `topic modeling' over time since 2018}
\label{fig.gtrend2}
\end{figure}

One should use, at most, a 90/10 sample split. As we have 241 observations, we only run Chow tests up to observation \#216, which corresponds to December 2021. For illustration purposes, Figure \ref{fig.gtrend2} plots the same data with the same scaling of the y-axis as shown in Figure \ref{fig.gtrend} but only for the period January 2018 - January 2024. One can already see that the `shock' observed in Figure \ref{fig.gtrend} seems less pronounced now. It is only due to a rescaling of the x-axis that emphasizes a significant point, which is statistics 101 knowledge, but often overlooked: Scaling matters. The often arbitrary choice of axis scaling is a strong argument for statistical testing for structural breaks. 

The grey shaded area in Figure \ref{fig.gtrend2} marks the time range for which we endogenously search for structural breaks. As said beforehand, December 2021 marks the end of this time frame, as we need sufficient observations after every hypothetical break. We begin with January 2020, as shortly afterwards, the first major decrease in searches for `topic modeling' occurred, as discussed beforehand. 

\begin{figure}[H]
\centering
\includegraphics[width = .8\linewidth]{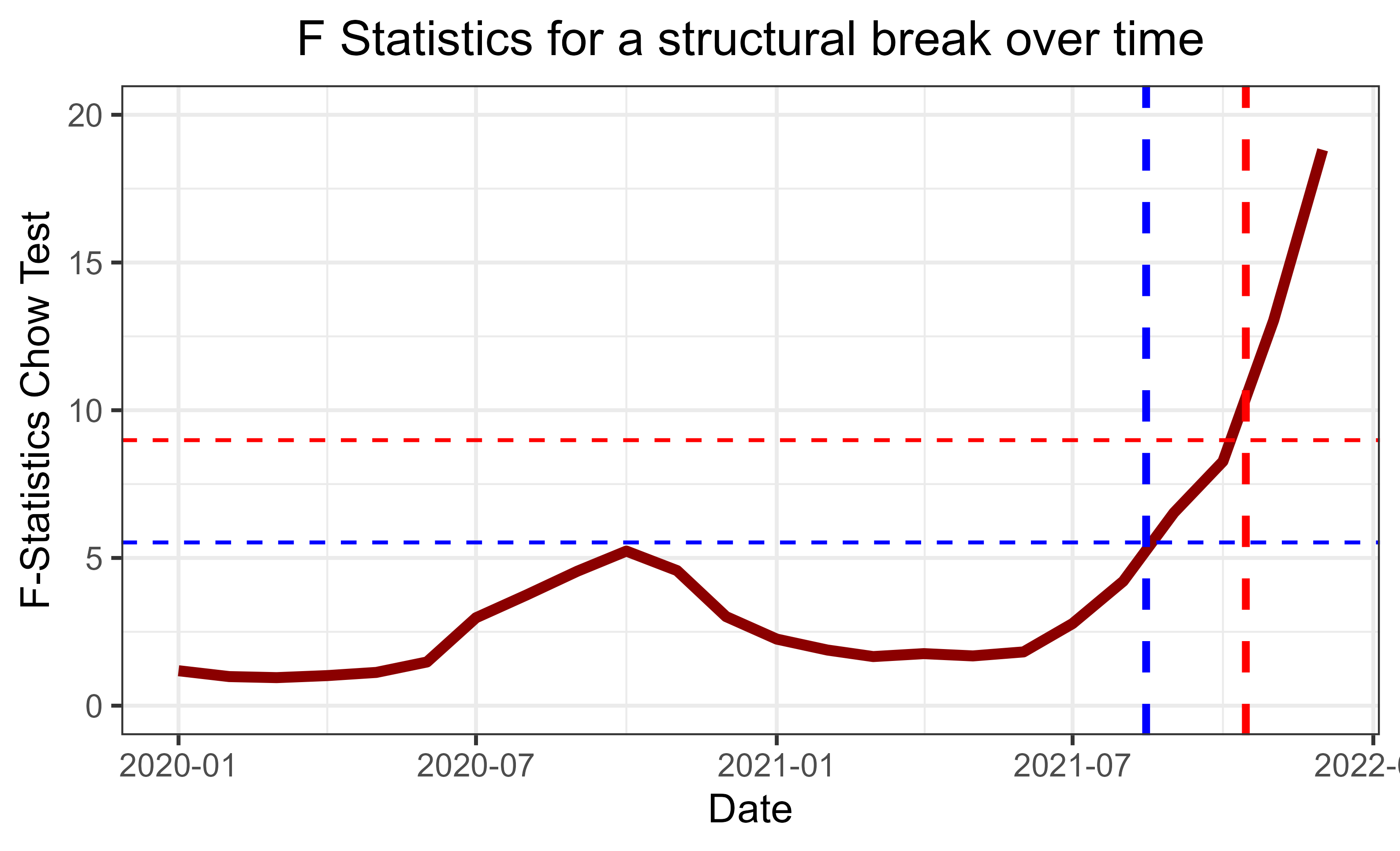}
\includegraphics[width = .8\linewidth]{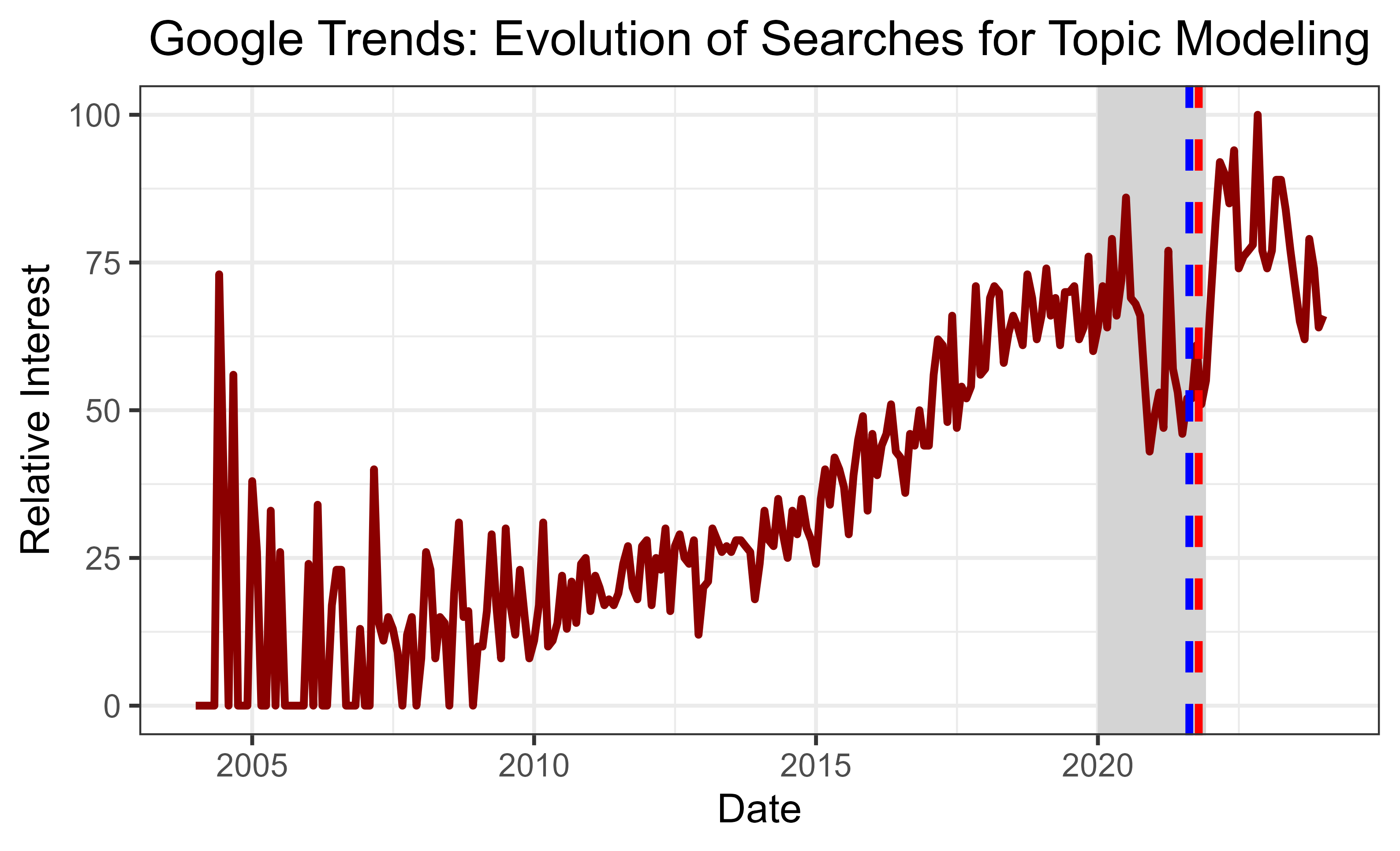}
\fnote{The upper panel displays the F statistics for the Chow testing procedure for endogenous breaks. The horizontal dashed lines represent the threshold values for which, with 95\% probability ($\alpha = 0.05$), a structural break exists. The red line displays the more conservative supremum criterium, and the blue line the average criterion as defined by \textcite{Andrews.1994}. The vertical dashed lines display the dates at which the structural breaks happen according to the statistical criteria. The lower panel depicts Figure \ref{fig.gtrend} again, but with added dashed lines for the statistically computed structural breaks as shown in the upper panel for the narrower time frame, which is marked with a grey shade in the lower panel.}
\caption{\normalsize F statistics for endogenous breakpoints}
\label{fig.fstats1}
\end{figure}

As sketched in the theoretical part of this chapter, we iteratively compute the F statistics for a structural break using the \verb|Fstats()| command. Figure \ref{fig.fstats1} presents these values using a dark red line. In addition, we have computed the boundary F value for both the average and the supremum criterion using separate specifications of the \verb|strucchange::boundary()| command for a 95\% confidence interval. This can be easily varied to a 90 or 99\% interval if necessary. For the `average' criterion, we observe for the dates from September 2021 onward that the F-statistic of the corresponding Chow test exceeds the minimum boundary. For the more conservative supremum criterium, the same holds for November and December 2021. For all earlier months since January 2020, the Chow test has not been able to reject the null of no break. But as stated beforehand, one must not confuse the first time the F statistic exceeds the boundary as evidence for a structural break at that point. 

So, let us go one step further and search for actual breakpoints or else confidence intervals in which breakpoints are most likely. 
The F statistic in the upper panel of Figure \ref{fig.fstats1} already hints at two areas:\footnote{Note that this is not evidence per se but is suggestive evidence where to look in more detail.} Late 2020 and from late 2021 onward again. It is backed by the visual inspection of the lower panel in Figure \ref{fig.fstats1}. In both cases, we observe a major downward shift. To look for the breakpoints, we can apply the powerful \verb|breakpoints()| command of the \verb|strucchange| package:
\begin{footnotesize}
\begin{verbatim}
break_p <- breakpoints(value ~ 1, h = 5, data = reduced_data)
confint(break_p)
Confidence intervals for breakpoints of optimal 4-segment partition: 
Call:
confint.breakpointsfull(object = break_p)
Breakpoints at observation number:
  2.5 % breakpoints 97.5 %
1     7          10     12
2    23          24     25
3    40          42     47
\end{verbatim}
\end{footnotesize}
The output is shortened as we want to focus on discrete breakpoints. As we look at the pure timeline, we regress the prevalence of our search term ``topic modeling'' only on a constant (1). As segment size, we set \verb|h=5|, because we use a reduced dataset that only starts in January 2020 and lasts until January 2024. This time frame covers 49 months, i.e., 10\% of it relates to $\approx$5 months. The size of the segment is, again, a rule of thumb and should be considered in the context of the respective dataset. With the second command \verb|confint()|, we can compute actual breakpoints, including a 95\% confidence interval. The breakpoints at \#10, \#24, and \#42 correspond to October 2020, December 2021, and June 2023. The first two dates are perfectly in line with the boundaries computed beforehand. The last date is too close to the ending of the time series and was, therefore, not considered in the computations shown in Figure \ref{fig.fstats1}. 

A closer visual inspection of the overall timeline in the lower panel of Figure \ref{fig.fstats1} reveals that a linear regression for the data points between late 2020 and late 2021 must have a negative slope. In contrast, the observations from 2022 on do not show a clear trend and suggest a relatively flat fitted line. It is in clear contrast to the steady growth trend since 2015. We can only speculate about the reason. More important is the insight that the Chow test for structural breaks has pointed out where the structural break has happened and that the likely COVID-19 shock in 2020 was just a setback but no structural break. The skeptical reader might respond that they would have seen the difference and there is nothing to debate. However, while one may argue about the visual shape of a line plot as shown in Figure \ref{fig.gtrend} shown beforehand, statistical tests are not subject to a verbal discussion on how to interpret a shape. The overall shape might not have suggested a break in the long-standing upward trend of search requests for `topic modeling.'

\FloatBarrier
\section{Conclusion}
\label{sec.conc}

This parsimonious practitioner's guide attempts to provide practical tips to researchers in the social sciences who work with topic models and compute expected topic prevalences. Two critical issues of time series are discussed: Non-stationarity and structural breaks. Both are of stellar importance in the applied work of social scientists as lots of research discusses either change over time or major disruptions. This paper can only touch upon the depths of the underlying econometrics and shall serve as an encouragement to engage further with time series econometrics. Three textbooks are cited, namely the work by \textcite{ Enders.2010, Hill.2008, Verbeek.2008}. The author's subjective criteria have selected them, as I find these three textbooks particularly helpful. However, many alternatives exist, and it is up to the reader to choose which book to read. 

Overall, this paper shall pave the way to more econometric rigor in studying topic models. In particular, it shall help overcome the reliance on only visual inspections. This guide shows that the coding effort in  \verb|R| is relatively low as the presented packages already provide the reader with ready-to-use results. Computing the suggested optimal lag length is also quite simple, as one only needs to insert the particular length $T$ of their time series into the presented rules of thumb. Hence, applying this small time series toolkit should hopefully be straightforward.

\subsubsection*{Data availability}
\begin{itemize}
\item The dataset based on Google Trends data is publicly available on \emph{Zenodo:} \url{https://doi.org/10.5281/zenodo.11047057}
\item Code files available on GitHub: \url{https://github.com/schmalwb/topic_model_time_series}. \emph{[note that the Python code is based on an R wrapper. Therefore, R needs to be installed. Detailed explanations provided in the code file]}
\end{itemize}


\newpage
\begin{singlespace}
\printbibliography
\addcontentsline{toc}{section}{\protect\numberline{}References}

\vspace{4mm}
\noindent \emph{\small The cited textbooks might have been published in newer editions already. Usually, updated versions can be used likewise as the discussed topics are unlikely to change substantially.}
\end{singlespace}
\newpage


\end{doublespace}
\end{document}